\documentclass[12pt]{article}
\usepackage{epsf,epsfig}

\newcommand{\be}{\begin{equation}}
\newcommand{\ee}{\end{equation}}
\newcommand{\ber}{\begin{eqnarray}}
\newcommand{\eer}{\end{eqnarray}}
\newcommand{\gsim}{\, \raisebox{-0.8ex}{$\stackrel{\textstyle >}{\sim}$ }}

\def\fm3{fm$^{-3}$}

\begin{document}
\baselineskip 16pt plus 2pt minus 2pt
\bibliographystyle{prsty2}

\begin{titlepage}

\par
\topmargin=-1cm      
{ \small

}

\vspace{20.0pt}

\begin{centering}

{\large \bf Chiral effective theory predictions for deuteron form
  factor ratios at  low $Q^2$}

\vspace{40.0pt} {\bf Daniel~R.~Phillips}\\
\vspace{20.0pt}
{\sl Department of Physics and Astronomy, Ohio
      University, Athens, OH 45701}\\
\vspace{2.0pt}
{\sl Department of Natural Sciences, Chubu University, 
1200 Matsumoto-cho, Kasugai, Aichi 487-8501, JAPAN}\\
\vspace{6.0pt}
{\tt Email: phillips@phy.ohiou.edu }\\

\vspace{15.0pt}
\end{centering}
\vspace{20.0pt}

\begin{abstract}
\noindent 
We use chiral effective theory ($\chi$ET) to predict the deuteron form factor
ratio $G_C/G_Q$ as well as ratios of deuteron to nucleon form
factors. These ratios are calculated to next-to-next-to-leading
order. At this order the chiral expansion for the $NN$ isoscalar
charge operator (including consistently calculated $1/M$ corrections)
is a parameter-free prediction of the effective theory. Use of this
operator in conjunction with NLO and NNLO $\chi$ET wave functions
produces results that are consistent with extant experimental data for
$Q^2 < 0.35$ GeV$^2$. These $\chi$ET wave functions predict a deuteron
quadrupole moment $G_Q(Q^2=0)=0.278$--$0.282~{\rm fm}^2$---with the
variation arising from short-distance contributions to this
quantity. The variation is of the same size as the discrepancy between
the theoretical result and the experimental value. This motivates the
renormalization of $G_Q$ via a two-nucleon operator that couples to
quadrupole photons.  After that renormalization we obtain a robust
prediction for the shape of $G_C/G_Q$ at $Q^2 < 0.3~{\rm GeV}^2$. This
allows us to make precise, model-independent predictions for the
values of this ratio that will be measured at the lower end of the
kinematic range explored at BLAST.  We also present results for the
ratio $G_C/G_M$.
\end{abstract}

\vspace*{10pt}
\begin{center}
\end{center}
\vfill

\end{titlepage}


\par
\topmargin=-1cm      

\vspace{20.0pt}

\section{Introduction}

\label{sec-intro}

The static electromagnetic properties of deuterium provide interesting
information on the dynamics at work within the nucleus. The fact that
deuterium's charge is one teaches us little other than the validity of
charge conservation in the nuclear system, but that its magnetic
moment $\mu_d \neq \mu_n + \mu_p$ and that it has of a non-zero
quadrupole moment  are facts which played an important role in
establishing that non-central components of the $NN$
potential are at work within the deuterium nucleus. 

The most accurate value for the deuteron quadrupole moment comes from
a  molecular physics experiment~\cite{CR71,BC79}. It is:
\begin{equation}
Q_d=0.2859(3)~{\rm fm}^2.
\label{eq:Qdexpt}
\end{equation}
Meanwhile the best determination of $\mu_d$ comes from the
spectroscopy of the deuterium atom. It is~\cite{CODATA}:
\begin{equation}
\mu_d=0.8574382284(94)~\mu_N.
\end{equation}
But elastic electron scattering from deuterium---provided the
one-photon-exchange approximation is valid---probes
M1 and E2 responses for
(virtual) photons that have a finite three-momentum ${\bf q}$. The
relevant form factors are related to Breit-frame matrix elements of
the two-nucleon four-current $J_\mu$ via
\begin{eqnarray}
G_M&=&-\frac{1}{\sqrt{2 \eta} |e|} 
\left \langle1\left|J^+\right|0\right \rangle,
\label{eq:GM}\\
G_Q&=&\frac{1}{2 |e|\eta M_d^2} 
\left(\left \langle 0\left|J^0\right|0 \right \rangle
- \left \langle 1\left|J^0\right|1 \right \rangle\right).
\label{eq:GQ}
\end{eqnarray}
These, together with the charge form factor, $G_C$:
\begin{equation}
G_C=\frac{1}{3 |e|} \left(\left \langle 1\left|J^0\right|1 \right \rangle + 
\left \langle 0\left|J^0\right|0 \right \rangle + \left \langle -1\left|J^0\right|-1 \right \rangle
\right),\label{eq:GC}
\end{equation}
provide a complete set of invariant functions for the description of
the deuterium four-current that interacts with the electron's current
in this approximation. In Eqs.~(\ref{eq:GM})--(\ref{eq:GQ}) 
we have labeled the deuteron states by the projection of the
deuteron spin along the direction of the three-vector ${\bf p}_e'
-{\bf p}_e \equiv {\bf q}$, and $\eta \equiv Q^2/(4 M_d^2)$, with
$Q^2=|{\bf q}|^2$ since we are in the Breit frame. We can then
calculate the deuteron structure functions:
\begin{eqnarray}
A&=&G_C^2 + \frac{2}{3} \eta G_M^2 + \frac{8}{9} \eta^2  M_d^4
G_Q^2,\\
\label{eq:A}
B&=&\frac{4}{3} \eta (1 + \eta) G_M^2.
\label{eq:B}
\end{eqnarray}
In terms of $A$ and $B$ the one-photon-exchange interaction yields a
lab. frame differential cross section for electron-deuteron scattering
\begin{equation}
\frac{d \sigma}{d \Omega}=\frac{d \sigma}{d \Omega}_{\rm NS}
\left[A(Q^2) + B(Q^2) \tan^2\left(\frac{\theta_e}{2}\right)\right],
\label{eq:dcs}
\end{equation}
Here $\theta_e$ is the electron scattering angle, and $\frac{d
  \sigma}{d \Omega}_{\rm NS}$ is the (one-photon-exchange) cross 
section for electron
scattering from a point particle of charge $|e|$ and mass $M_d$.  The
form factors defined in Eqs.~(\ref{eq:GM})--(\ref{eq:GC}) are related
to the static moments of the nucleus by:
\begin{eqnarray}
G_C(0)&=&1,\\
G_Q(0)&=&Q_d,\\
G_M(0)&=&\mu_d \frac{M_d}{M},
\end{eqnarray}
with $M$ the nucleon mass.
For recent reviews of experimental and theoretical work on
elastic electron-deuteron scattering see Refs.~\cite{GG01,vOG01,Si01}.

From Eq.~(\ref{eq:dcs}) it is already clear that measurements of the
differential cross section alone cannot yield uncorrelated information
on all three form factors. To measure $G_Q$, $G_M$, and $G_C$ in a
model-independent way one must obtain data with polarized deuterium
targets or polarized electron beams. A new set of measurements of
polarization observables in electron-deuteron scattering will soon be
available from the data set obtained at the Bates Large-Acceptance
Spectrometer Toroid (BLAST). There polarized electrons of energy 850
MeV circulated in the Bates ring and were scattered from an internal
target containing polarized deuterium.  The significant amount of beam
on target (3 million Coulombs since late 2003), and high degree of
beam and target
polarization achieved at BLAST, means that we anticipate data on
electron-deuteron polarization observables that is more precise than
that derived from any previous measurement.

The electron beam circulating in the Bates ring was roughly 70\%
polarized, and the deuterium target employed could operate in both a
vector-polarized and tensor-polarized mode. This gives access to all
of the elastic electron scattering deuterium structure
functions. Prominent among these are $t_{11}$, the vector analyzing
power, and $t_{20}$, the tensor analyzing power. They are related to
the form factors defined above by~\cite{vOG01}
\begin{eqnarray}
t_{20}=-\sqrt{2} \frac{x(x + 2) + \frac{1}{2} y^2 (1 + 2 \varepsilon)}{1 +
  2 (x^2 + y^2(1 + 2 \varepsilon))}; \quad
t_{11}=2 \sqrt{\varepsilon} y \frac{1 + \frac{x}{2}}{1 + 2(x^2   + y^2 (1 + 2 \varepsilon))},
\end{eqnarray}
where the ratios $x$ and $y$ are:
\begin{eqnarray}
x &=& \frac{2}{3} \, \eta \, \frac{G_Q M_d^2}{G_C},\\
y&=& \sqrt{\frac{\eta}{3}} \frac{G_M}{G_C},
\end{eqnarray}
and $\varepsilon=(1 + \eta) \tan^2\left(\frac{\theta_e}{2}\right)$.
Therefore measurements of $t_{11}$ and $t_{20}$ should facilitate
the extraction of the ratios $G_M/G_C$ (which at the
$Q^2$'s we will consider here mainly affects $t_{11}$) and $G_Q/G_C$
(which mainly affects $t_{20}$). In this paper we provide
predictions for these ratios which are based on chiral effective
theory ($\chi$ET).

This approach (for reviews see Refs.~\cite{Be00,BvK02,Ph05,Ep06}) is
based on the use of a chiral expansion for the physics of the
two-nucleon system. In the formulation suggested by
Weinberg~\cite{We90,We91,We92}, the $\chi$ET treatment of the $NN$
system is based on a systematic chiral and momentum expansion for the
two-nucleon-irreducible kernels of the processes of interest. In
particular, wave
functions are computed using an $NN$ potential expanded up to a given
order in the small parameter:
\begin{equation}
P \equiv \frac{p,m_\pi}{\Lambda}
\label{eq:P}
\end{equation}
where $p$ is the momentum of the nucleons and $\Lambda$ is the
breakdown scale of the theory. For electron-deuteron scattering the
other two-nucleon-irreducible kernel that must be calculated is the
deuteron current operator $J_\mu$. We also expand this object as:
\begin{equation}
J_\mu=e \sum_{i=1}^\infty \xi_i \frac{1}{\Lambda^{i-1}}{\cal O}_\mu^{(i)},
\label{eq:sum}
\end{equation}
where the operator ${\cal O}_\mu^{(i)}$ contains $i-1$ powers of the
small parameter $P$, which now includes the momentum transfer to the
nucleus, $q$, as one of the small scales in the
numerator. For chiral effective theories without
an explicit Delta degree of freedom $\Lambda$ will in general be
$M_\Delta - M$, but in electron-deuteron elastic scattering the
$\Delta N$ intermediate state is not allowed and so $\Lambda$ will be
larger, $\Lambda \gsim 2(M_\Delta - M)$. 

The $NN$ potential has now been computed up to
$O(P^2)$~\cite{Or96,Ep99}, $O(P^3)$~\cite{Or96,Ep99,EM01} and
$O(P^4)$~\cite{EM03,Ep05}. In this paper we will employ wave functions
computed using the next-to-leading order [NLO=$O(P^2)$],
next-to-next-to-leading order [NNLO=$O(P^3)$] and N$^3$LO
potentials [$O(P^4)$] developed in Ref.~\cite{Ep05}. These potentials
are regularized in two different ways: first, spectral-function
regularization (SFR) at a scale $\bar{\Lambda}$~\cite{Ep04}, is
applied to the two-pion contributions. Then, after the SFR potential
$V_{ll'}^{sj}(p,p')$ is obtained in a particular $NN$ partial wave, it
is multiplied by a regulator function $f$, so that the
Lippmann-Schwinger equation can be straightforwardly solved:
\begin{equation}
V_{ll'}^{sj}(p,p') \rightarrow 
f(p) V_{ll'}^{sj}(p,p') f(p') \quad \mbox{with} \quad
f(p)=\exp\left(-\frac{p^6}{\Lambda^6}\right).
\label{eq:NNregulator}
\end{equation}

There have been questions raised as to the consistency of the wave
functions computed in this way~\cite{Ka96,Be01,ES02,No05,PVRA05,Bi05}.  Partly
because of these questions we will, for comparison, also present
results for electron-deuteron matrix elements using the form
(\ref{eq:sum}) for the current operator, and wave functions derived
from the Nijm93~\cite{St94} or CD-Bonn~\cite{Ma01} potentials, as well
as potentials with one-pion exchange at long range and a square well
and surface delta function of radius $R$~\cite{PC99}. We stress that
such calculations are not chirally consistent. However, common
features of deuteron observables that can be identified within
calculations that use these different types of wave functions---chiral
effective theory, potential models, and one-pion-exchange
tails---should be independent of the details of physics at ranges $\ll
1/m_\pi$ in deuterium, and so should not be sensitive to any
subtleties pertaining to the renormalization of the $\chi$ET.

The operators ${\cal O}_\mu^{(i)}$ and the coefficients $\xi_i$ in
Eq.~(\ref{eq:sum}) are constructed according to the counting rules and
Lagrangian of heavy-baryon chiral perturbation theory (HB$\chi$PT),
which is reviewed in Ref.~\cite{Br95}. Here the results we will
present for $G_C$ and $G_Q$ include all contributions to $J_\mu$ up to
chiral order $eP^3$. This is the next-to-next-to-leading order (NNLO)
for these quantities.  Calculations of electron-deuteron scattering
with the NNLO $\chi$ET operator were already considered in
Ref.~\cite{Ph03}, which improved upon results with the $O(eP^2)$
operator in Ref.~\cite{WM01} and the $O(e)$ results of
Ref.~\cite{PC99}.  However, as was already observed in
Ref.~\cite{Ph03} and is reiterated below, calculation of the
quadrupole combination of matrix elements at NNLO does not reproduce
the experimental value of $Q_d$ to the accuracy one would expect at
that order. We identify the cause of this as short-distance two-body
contributions to $J_0$ of natural size (i.e. with $\xi_i \sim 1$)
through which quadrupole photons induce an $L=0 \rightarrow L=0$
transition in the $S=1$ deuteron state~\cite{PC99,Ch99}.  We use the
operator induced by these short-distance contributions to renormalize
the deuteron quadrupole moment, and hence the form factor $G_Q$.  We
also provide results for $G_M$ up to NLO. Not surprisingly, $G_M$ at
NLO proves more sensitive to short-distance physics than does the
renormalized $G_Q$.

Throughout this work we will use the factorization of nucleon
structure employed in Ref.~\cite{Ph03} in order to include the effects
of finite nucleon size in the calculation. There it was shown that the
chiral expansion for the ratios:
\begin{equation}
\frac{G_C}{G_E^{(s)}} \quad \mbox{and} \quad \frac{G_Q}{G_E^{(s)}} \quad
\mbox{and} \quad \frac{G_M}{G_M^{(s)}},
\label{eq:ratios}
\end{equation}
with $G_E^{(s)}$ and $G_M^{(s)}$ the isoscalar single-nucleon electric
and magnetic form factors, is better behaved than the chiral expansion
for $G_C$, $G_Q$, and $G_M$ themselves.  The ratios (\ref{eq:ratios})
allow us to focus on the ability of the chiral expansion to describe
deuteron structure, and we will employ the $\chi$ET results for the
ratios in our efforts to predict BLAST's results for polarized
electron scattering from a deuterium target. We note that, up to the
order we work to here, our predictions for $G_C/G_Q$ are independent
of the manner in which we include nucleon structure in the
calculation. Our invoking the factorization of nucleon structure in
the electron-deuteron matrix elements plays no role in our predictions
for $G_C/G_Q$.

The chiral perturbation theory for this calculation was laid out in
Refs.~\cite{Ph03,WM01}, and so here we merely summarize the pertinent
features of the chiral expansion for the deuteron currents in
Section~\ref{sec-kernel}. However, in doing so we find that we must
address the issue of how to calculate the corrections to this ratio
that have coefficients which are fixed by low-energy Lorentz
invariance. We deal with this problem in Section~\ref{sec-oneoverM},
by recalling results of Friar, Adam {\it et al.}, and Arenh\"ovel {\it
  et al.}, which show that such corrections can be calculated
unambiguously, as long as they are included consistently in both the
$NN$ potential and the current operator.  Then, in
Section~\ref{sec-other} we discuss effects in the $J_0$ operator
beyond $O(e P^3)$, and explain how we will estimate their impact on
$G_C$ and $G_Q$.  In particular, we write down an operator that
represents the effects of physics at mass scales above 1 GeV on $G_Q$,
and can repair the discrepancy between the experimental value of $Q_d$
and our predictions for $G_Q(0)$. We also discuss how to estimate the
remaining uncertainty in our results.  Then, in
Section~\ref{sec-J0results} we present results for $G_C$, $G_Q$, and
the ratio $G_C/G_Q$.  We show that the shape of $G_Q$ can be predicted
in a model-independent way for $Q^2 < 0.3$ GeV$^2$, but the
uncertainty in the ratio $G_C/G_Q$ is sizeable at $Q^2 \approx 0.4$
GeV$^2$.  Finally, in Section~\ref{sec-Jplusresults} we present
results for $G_C/G_M$, and in Section~\ref{sec-conclusion} we
summarize and provide an outlook.

\section{The deuteron current}

\label{sec-kernel}

We now discuss the charge and current operators in turn.  Such a
decomposition is, of course, not Lorentz invariant, so here we make
this specification in the Breit frame. 

\subsection{Deuteron charge}
\begin{figure}[htbp]
\vspace{0.2cm}
\hspace{-0.25in}
\centerline{\epsfig{figure=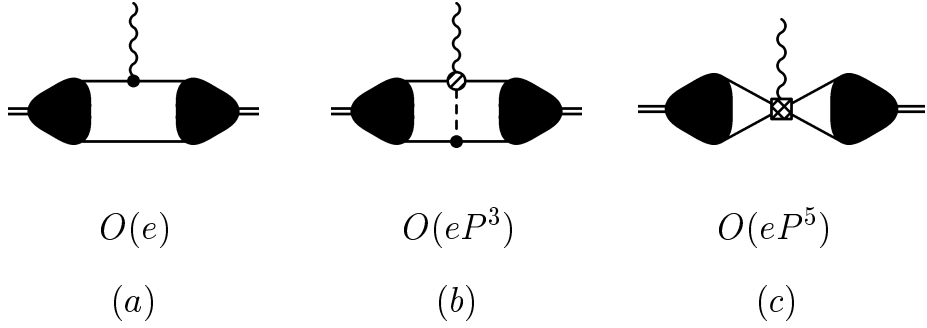,width=11.0cm}}
\vspace{-0.1cm}
\caption{Diagrams representing the leading contribution
to the deuteron charge operator [(a)], the leading two-body contribution 
to $J_0$ [(b)], and the dominant short-distance piece [(c)].
Solid circles are vertices from ${\cal L}_{\pi N}^{(1)}$, and the shaded
circle is the vertex from ${\cal L}_{\gamma \pi N}^{(2)}$.}
\label{fig-twobodycharge}
\end{figure}

The vertex from ${\cal L}_{\pi N}^{(1)}$ which represents an $A_0$ photon
coupling to a point nucleon gives the leading-order (LO) contribution to $J_0$
as depicted in Fig.~\ref{fig-twobodycharge}(a). At $O(e P^2)$ this is
corrected by insertions in
${\cal L}_{\pi N}^{(3)}$ that generate the nucleon's isoscalar charge
radius. This gives a result for $J_0$ through $O(eP^2)$:
\begin{equation}
J_0^{(2)}=|e| \left(1 - \frac{1}{6} \langle r_{Es}^2 \rangle
Q^2\right) + j_0^{(1/M^2)}({\bf q})
\label{eq:structure}
\end{equation}
with $j_0^{(1/M^2)}({\bf q})$ the
``relativistic'' corrections to the single-nucleon charge
operator. These contributions have fixed coefficients that are
determined by the requirements of Poincar\'e invariance. Since these
coefficients scale as $1/M^2$ this particular set of $O(e P^2)$
contributions are generally smaller than one would estimate given the
formula (\ref{eq:P}) for the parameter $P$. These ``relativistic''
corrections can be calculated by writing down a $J_0$ operator that,
when inserted between deuteron wave functions calculated in the
two-nucleon center-of-mass frame, yields results for the matrix
elements that are Lorentz covariant up to the order to which we
work. To do this we employ the formalism of Adam and Arenh\"ovel, as
described in Ref.~\cite{AA96}. 

Meanwhile the only contribution at $O(eP^3)$, or NNLO, comes from the
tree-level two-body graph shown in Fig.~\ref{fig-twobodycharge}(b).
In HB$\chi$PT the relevant single-nucleon photo-pion vertex arises as
a consequence of the Foldy-Wouthuysen transformation which generates a
term in ${\cal L}^{(2)}$.  Straightforward application of the Feynman
rules for the relevant pieces of the HB$\chi$PT Lagrangian gives the
result for this piece of the deuteron current~\cite{Ph03}:
\begin{equation}
\langle {\bf p}'|J_0^{(3)}({\bf q})|{\bf p} \rangle=
\tau_1^a \tau_2^a \, \, \frac{|e| g_A^2}{8 f_\pi^2 M}  \, \,  
\left[\frac{\sigma_1 \cdot {\bf q} \, \sigma_{2} \cdot 
({\bf p} - {\bf p}' + {\bf q}/2)}
{m_\pi^2 + ({\bf p} - {\bf p}' + {\bf q}/2)^2} + (1 \leftrightarrow 2)\right],
\label{eq:J02B}
\end{equation}
where ${\bf p}$ and ${\bf p}'$ are the (Breit-frame) relative momenta
of the two nucleons in the initial and final-state
respectively (see Ref.~\cite{Ri84} for a much earlier derivation).

Thus we now have a result for the deuteron's charge operator which can
be summarized as:
\begin{equation}
\langle {\bf p}'|J_0({\bf q})|{\bf p}\rangle=
\left[|e| \left(1 - \frac{1}{6} \langle r_{Es}^2 \rangle Q^2\right)
+ j_0^{(1/M^2)}({\bf q})\right] \delta^{(3)}(p' - p - q/2)
+ \langle {\bf p}'|J_0^{(3)}({\bf q})|{\bf p} \rangle
+ O(eP^4).
\label{eq:pure}
\end{equation}
However, it was shown in Ref.~\cite{Ph03} that the parameterization
(\ref{eq:structure}) of the nucleon's isoscalar charge distribution
breaks down at $|{\bf q}| \approx 300$ MeV. In order to avoid this difficulty
we observe that the result (\ref{eq:pure}) may be recast, up to the
order to which we work, as:
\begin{equation}
\langle {\bf p}'|J_0({\bf q})|{\bf p}\rangle=
\left[\left(|e| + j_0^{(1/M^2)}({\bf q})\right) \delta^{(3)}(p' - p - q/2) 
 + \langle {\bf p}'|J_0^{(3)}({\bf q})|{\bf p} \rangle\right] G_E^{(s)}(Q^2)
+ O(eP^4),
\label{eq:factor}
\end{equation}
with $G_E^{(s)}(Q^2)$ the complete one-loop HB$\chi$PT result for the
nucleon's
isoscalar electric form factor~\cite{He98}. This means that we can
write a result that is independent of $\chi$PT's difficulties in
describing nucleon structure if we focus on the ratio of $\langle {\bf p}'|J_0({\bf
  q})|{\bf p}\rangle$ to $G_E^{(s)}(Q^2)$.  We will then use experimental
data~\footnote{There is, of course, some circularity here, since
  electron-deuteron elastic scattering data has been used to constrain
  the neutron electric form factor, see, in particular,
  Ref.~\cite{Pl90}.}, in particular the parameterization of
Mergell {\it et al.}~\cite{Me95}, for $G_E^{(s)}(Q^2)$ in all computations we
present below. This use of experimental data for the single-nucleon
matrix element that appears in Eq.~(\ref{eq:factor}) allows us to
focus on how well the $\chi$ET is describing deuteron
structure, since it removes the nucleon-structure issues from the
computation of the deuterium matrix element. Our technique to achieve this
is rigorous, up to the chiral order to which we work here. 
The ratio $G_C/G_Q$ can also be computed independent of
nucleon-structure issues, as is made clear by a brief examination of 
Eq.~(\ref{eq:factor}), together with the definitions (\ref{eq:GC}) and (\ref{eq:GQ}).

\subsection{Deuteron three-current} 
The counting for the isoscalar three-vector current ${\bf J}$ was
already considered in detail by Park and
collaborators~\cite{Pa95}. ${\bf J}$ begins at $O(eP)$, 
but at $O(e P^3)$ there are finite-size and relativistic
corrections, which are suppressed by two powers of
$P^2$. This is the highest order we will
calculate $G_M$ to here, and at this order, using factorization we have:
\begin{equation}
\langle {\bf p}'|{\bf J}^{(3)}|{\bf p} \rangle=[|e| ({\bf p} + {\bf q}/2)/M + i \mu_S {\bf \sigma}
\times {\bf q} + {\bf J}^{(1/M^2)}({\bf q})] G_M^{(s)}(Q^2)
\delta^{(3)}(p' - p - q/2).
\end{equation}
with ${\bf p} - {\bf q}/4$ is the momentum of the struck nucleon, and
$\mu_S$ is the isoscalar magnetic moment of the nucleon, whose value we
take to be $\mu_S= 0.88 |e|/(2M)$.

At $O(e P^4)$ [NNLO] two kinds of magnetic two-body current enter the
calculation. One is short-ranged, and one is of pion
range~\cite{Pa95,Ch99,Sc99,Pa00,WM01}. Each of them has an
undetermined coefficient.  In principle those coefficients should be
fit to data (e.g. the deuteron magnetic moment, which is not exactly
reproduced by the current ${\bf J}$ and the wave functions employed
here) and the low-$Q^2$ shape of the form factor.

\section{Unitary equivalence and the consistent treatment of $1/M$ 
corrections}

\label{sec-oneoverM}

In this section we discuss the constraints imposed by Poincar\'e
invariance---or the low-energy manifestations thereof---on the
Breit-frame isoscalar $NN$ charge operator. Recall that in HB$\chi$PT
$\langle {\bf p}'|J_0^{(3)}({\bf q})|{\bf p} \rangle$ arises from a
piece of ${\cal L}_{\pi N}^{(2)}$ that has a fixed coefficient
obtained via a Foldy-Wouthuysen transformation. Therefore this piece of
the charge operator is a low-energy consequence of Lorentz covariance
of $\langle M'|J_\mu|M \rangle$. As such the contribution
(\ref{eq:J02B}) should be computed in a manner consistent with that
used to derive the $1/M^2$ corrections to the one-pion exchange part
of the $NN$ potential. Those $1/M^2$ corrections can be obtained from
the chiral Lagrangian--- specifically from the $1/M$ pieces in ${\cal
  L}_{\pi N}^{(2)}$ and the $1/M^2$ pieces in ${\cal L}_{\pi
  N}^{(3)}$. But the relevant operators involve the energy of the
individual nucleons, and so it is not immediately obvious how to
convert them to contributions to an energy-independent
quantum-mechanical potential. In fact, in the 1970s and 1980s many
techniques were developed by which quantum-mechanical operators could
be obtained from a relativistic quantum field
theory~\cite{AA96,Ad93,Fr80}. In all of these techniques there was
freedom in choosing whether (and if so, which) nucleon lines to put on
shell, as well as freedom in how to include meson retardation. As we
shall see, the choices made with respect to these two issues have an
impact on the form of the operators (both $V$ and $J_0$) that
are obtained.  Ultimately though, as long as operators and potentials
are derived in a consistent way, the different choices are related by
unitary transformations that leave matrix elements
unaffected~\cite{Ad93,Fr80,Fo99}.

That unitary transformation is
labeled by two parameters: $\nu$, which parameterizes the energy,
$k_0$, of the exchanged pion via
\begin{equation}
k_0^2=\left(1 - 2 \nu\right) \frac{(p'^2 - p^2)^2}{4 M^2},
\end{equation}
where $p'$ ($p$) is the length of the relative three-momentum vector
of the $NN$ system after (before) the meson exchange; and
$\beta$, which denotes a choice for the change in the nucleons'
energy after absorption of the pion~\cite{Ad93}:
\begin{equation}
(p_0' - p_0)_{1,2}=(1 - 2 \beta) \frac{p'^2 - p^2}{2 M}.
\end{equation}
Note that in quantum mechanics energy is not conserved at each vertex,
and so $(p_0' - p_0)_1=k_0=-(p_0'-p_0)_2$ need not necessarily hold.
Indeed, it turns out that since we are in the $NN$ c.m. frame the 
difference $p_0' - p_0$ is the same for both nucleons once an energy shell is
chosen.  The full expression for the $1/M^2$ corrections to the
one-pion-exchange potential in the case of arbitrary $\beta$ and $\nu$
can be found in Ref.~\cite{Ad93}. The main result for our purposes
here is that if $\beta=\frac{1}{4}$ then the potential takes the form:
\begin{equation}
\langle {\bf p}'|V^{(1 \pi)}|{\bf p} \rangle=-\tau_1^a \tau_2^a \frac{g_A^2}{4 f_\pi^2}
\frac{\sigma_1 \cdot ({\bf p}' - {\bf p})  \sigma_2 \cdot ({\bf p}' -
  {\bf p})}{({\bf p}' - {\bf p})^2 + m_\pi^2} \left(1 - \frac{p'^2 +
  p^2}{2 M^2} + O\left(\frac{1}{M^4}\right)\right).
\label{eq:relOPE}
\end{equation}
This is the one-pion-exchange potential used in the N$^3$LO
computation of Ref.~\cite{Ep05}. (Corrections to ``leading'' two-pion
exchange diagrams that are suppressed by $1/M$ are also included
there, but are associated with pieces of the $J_0$ which are of higher
order than we work to here.) The computation of
Ref.~\cite{Ma01} employed the form for $\langle {\bf p}'|V^{(1
  \pi)}|{\bf p} \rangle$ that corresponds to $\beta=0$. All other
potentials we discuss here (including the NNLO and NLO ones used in
Ref.~\cite{Ep05}) employed the non-relativistic form of OPE, i.e. the
result (\ref{eq:relOPE}), but without the additional factor in the
round brackets. Meanwhile, all of the potentials we have used neglect
retardation, which means they have set $\nu=\frac{1}{2}
\Leftrightarrow k_0=0$.

Consistent reduction of the contributions to the deuteron charge
operator then leads to a more general result for diagram
Fig.~\ref{fig-twobodycharge}(b) than that given in
Eq.~(\ref{eq:J02B})~\cite{Ad93}:
\begin{eqnarray}
&& \langle {\bf p}'|J_0^{(3)}({\bf q})|{\bf p} \rangle=
\tau_1^a \tau_2^a \, \, \frac{|e| g_A^2}{8 f_\pi^2 M}  \, \,  
\left[(1-\beta) \frac{\sigma_1 \cdot {\bf q} \, \sigma_{2} \cdot 
({\bf p} - {\bf p}' + {\bf q}/2)}
{m_\pi^2 + ({\bf p} - {\bf p}' + {\bf q}/2)^2} 
\right.\nonumber\\
&& \qquad - 
\frac{1 - \nu}{2}
\left.\frac{\sigma_1 \cdot ({\bf p} - {\bf p}' + {\bf q}/2) \, \sigma_{2} \cdot 
({\bf p} - {\bf p}' + {\bf q}/2) \, \, \, {\bf q} \cdot ({\bf p} - {\bf p}' + {\bf q}/2)}
{[m_\pi^2 + ({\bf p} - {\bf p}' + {\bf q}/2)^2]^2} + (1 \leftrightarrow
2)\right].\nonumber\\
\label{eq:J02Bmutilde}
\end{eqnarray}

In Eq.~(\ref{eq:J02B}) we obtained the result for the $O(e P^3)$ piece
of the charge operator that corresponds to $\beta=0$ and $\nu=1$,
because the field-theoretic manipulations used to arrive at
Eq.~(\ref{eq:J02B}) assume that the fields represent physical
particles, i.~e. they are on-shell. The result (\ref{eq:J02Bmutilde})
may be obtained from Eq.~(\ref{eq:J02B}) by applying a unitary
transformation~\cite{Ad93,Fr80}:
\begin{equation}
J_0^{(3)}(\beta,\nu)=U^\dagger(\beta,\nu) J_0^{(3)}(0,1) U(\beta,\nu),
\label{eq:J0unitary}
\end{equation}
where the form of $U$ can be found in the original papers.
The same unitary transformation generates consistent
$1/M^2$ corrections to the one-pion-exchange part of the $NN$
potential:
\begin{equation}
V_{\rm OPE}(\beta,\nu)=U^\dagger(\beta,\nu) V_{\rm OPE}(0,1) U(\beta,\nu),
\label{eq:VOPEunitary}
\end{equation}
including the form (\ref{eq:relOPE}) if the choice
$\beta=\frac{1}{4}$, $\nu=\frac{1}{2}$ is adopted. 
This is not consistent with the choice made in obtaining
Eq.~(\ref{eq:J02B}) in Ref.~\cite{Ph03} because
the $NN$ potential of Ref.~\cite{Ep05} was 
computed using the Okubo formalism developed in Ref.~\cite{Ep99A}.

In Ref.~\cite{Ep05} this issue of the choice made for $\beta$ and
$\nu$ does not arise until the N$^3$LO potential is derived, because
in that paper, and in the earlier Refs.~\cite{Ep99A,Ep99}, Epelbaum
{\it et al.} chose to count $M \sim \Lambda^2$. In doing this they
were adhering to Weinberg's original argument as to why it is the
two-nucleon-irreducible kernel---and not the $NN$ amplitude
itself---which admits a chiral expansion. $NN$ intermediate states
introduce factors of $M$ in the amplitude for loop graphs, and if $M
\sim \Lambda^2$ then the $n$th iterate of the one-pion-exchange
potential is the same order as one-pion exchange
itself~\cite{We90,We91}. However, in Ref.~\cite{Be01} the need to
iterate the one-pion-exchange potential to all orders was established
without any reference to counting $M \sim \Lambda^2$, being justified
instead by the singular, and attractive, nature of the $NN$ force (see
also Ref.~\cite{Bi05}).  Therefore, while it is true that corrections
to the $NN$ potential which are suppressed by powers of $1/M$ are often
smaller than, e.g.  those arising from excitation of the Delta(1232),
in discussing electron-deuteron scattering we will consider a regime
in which $q/M$ can be sizeable.  Therefore we follow the original
HB$\chi$PT counting and take $M \sim \Lambda$. As we shall see, this
counting is supported by the fact that the contribution
(\ref{eq:J02B}) plays a significant, but not excessive, role in the
deuteron charge and quadrupole form factors.

If we count $M \sim \Lambda$ the dilemma presented by the
inconsistency between $V$ and $J_0$ arises already at $O(e P^3)$. A way out
of this dilemma is provided by Eqs.~(\ref{eq:J0unitary}) and
(\ref{eq:VOPEunitary}). They guarantee that we will obtain the same
result (up to $O(p^4/M^4)$ corrections) for deuteron matrix elements
$\langle M'|J_0|M \rangle$, regardless of what choices for $\beta$ and
$\nu$ we make when constructing the
operators $V$ and $J_0$ from the chiral effective field theory,
provided that we consistently include the $O(p^2/M^2)$
pieces of the potential $V$ and the $O(eP^3)$ pieces of the operator
$J_0$.  Therefore in order to be consistent with the calculation of
the $1/M^2$ corrections to $V_{\rm OPE}$ in Ref.~\cite{Ep05} we must
adopt the choice $\beta=\frac{1}{4}$ in the formula
(\ref{eq:J02Bmutilde}) for $J_0^{(3)}$. If we do this, and also make
sure to calculate one-pion exchange according to the formula
(\ref{eq:relOPE}), then our results for matrix elements of the
deuteron charge operator will incorporate the low-energy consequences
of Lorentz invariance, up to corrections of $O(p^4/M^4)$ (higher order
than we consider here). Note that the CD-Bonn potential is a different
case, since the use of a pseudoscalar $\pi NN$ coupling means that
there we have $\beta=0$. Therefore in that case, and only in that
case, we have used the expression (\ref{eq:J02B}) for the first part
of $J_0^{(3)}$, with no modification by the factor of $\frac{3}{4}$
that must be present if $V^{(1 \pi)}$ is constructed with $\beta=\frac{1}{4}$.

This still leaves us with the issue of how the $p^2/M^2$ corrections
in the one-pion exchange potential (\ref{eq:relOPE}) and the $p^2/M^2$
corrections to the nucleon kinetic energy operator are to be accounted
for in the calculations using the NNLO and NLO wave functions of
Ref.~\cite{Ep05} (or included in calculations with the Nijm93 wave
function of Ref.~\cite{St94} or the wave functions of
Ref.~\cite{PC99}). The original calculations of these wave functions
did not include such effects, but since we count $M \sim \Lambda$
here, we need to include them in order to have a consistent
calculation of $G_C$ and $G_Q$ to $O(e P^3)$.

Starting from the Kamada-Gl\"ockle transformation~\cite{KG98}, we show
in Appendix~\ref{ap-p2overM2corrns} the major part of these effects
can be included by making changes to the short-distance pieces of the
$NN$ potential, and using a slightly modified wave function $\psi$ in
the computation of $G_C$, $G_Q$, and $G_M$. That wave function is
related to the original non-relativistic wave function
$\tilde{\psi}$ obtained in Ref.~\cite{Ep05} by~\cite{KG98}:
\begin{equation}
\psi(p)=\sqrt{\frac{M}{\sqrt{M^2 + p^2}}} \left(\frac{2M}{M + \sqrt{M^2
      + p^2}}\right)^{1/4}
\tilde{\psi}\left(\sqrt{2M \sqrt{M^2 + p^2} - 2M^2}\right).
\label{eq:wfreln2}
\end{equation}
The solution of the non-relativistic Schr\"odinger equation for the
wave function $\tilde{\psi}$, followed by the use of the formula
(\ref{eq:wfreln2}) to obtain the solution of the relativistic
Schr\"odinger equation, is the method by which we incorporate $1/M^2$
effects for the NLO and NNLO wave functions of Ref.~\cite{Ep05}, the
wave function of Ref.~\cite{St94}, and the wave functions of
Ref.~\cite{PC99}.
The effects of using $\psi$, rather
than the wave function, $\tilde{\psi}$, to calculate electron-deuteron
observables increase with photon momentum transfer $|{\bf q}|$, but are small
over the entire range for which the $\chi$ET predictions can be
trusted. At $|{\bf q}|=700$ MeV for the Nijm93 wave function they change $G_C$
by 6.3\%, $G_Q$ by 1.2\%, and $G_M$ by 2.0\%. (The effect on $G_C$ is
proportionately larger because 700 MeV is quite close to the
form factor minimum.)

\section{Short-distance $NN$ charge operators and $Q_d$}

\label{sec-other}

So far we have obtained the deuteron two-body charge operator up to
$O(eP^3)$, or next-to-next-to-leading order. This is the order up to
which the calculation we present here is fully systematic. In this
section we discuss the role of contributions that are nominally higher
order, and identify one particular mechanism that apparently could
generate significant effects at $Q^2=0$. We are particularly
interested in this operator because ``the $Q_d$ problem'' that is
present in all modern potential models (see, e.g Ref.~\cite{St94,Wi95})
persists in the $\chi$ET. The problem is that all such calculations
under-predict the value (\ref{eq:Qdexpt}) for $Q_d$ by about 2--3\% when
they use a charge operator that includes all effects up to NNLO in the
$\chi$ET. The remaining discrepancy is large compared to the expected
$P^4$ size of higher-order effects. It is also large compared to other
discrepancies between theory and experiment in the
${}^3$S$_1$-${}^3$D$_1$ channel of the $NN$ system.

And the situation is actually worse than this, because at $O(e
P^4)$---one order higher than we are considering here---there are
two-meson-exchange contributions to the deuteron charge
operator. One might hope that these provide the missing strength in
the E2 response of deuterium at $Q^2=0$.
These diagrams are presently being calculated for finite $Q^2$, and
will be
incorporated in a future computation of the charge and quadrupole form
factors~\cite{quadri}. However, it is already known that they do not
give a sizeable contribution to the deuteron quadrupole moment. Park
{\it et al.}~\cite{Pa00} computed their effect on $Q_d$ using the AV18
wave function~\cite{Wi95}, and found:
\begin{equation}
\Delta Q_d^{(4)}=-0.002~{\rm fm}^2.
\end{equation}
Therefore these effects will {\it not} repair the discrepancy between
the calculated $G_Q(0)$ and the experimental $Q_d$.

At the next order, $O(e P^5)$, there are additional two-pion-exchange
contributions to the deuteron charge. However, short-distance two-body
currents that contribute to $\langle r_d^2 \rangle$ and $Q_d$ are also
present, and are depicted in Fig.~\ref{fig-twobodycharge}(c).  Even
though it is suppressed by five powers of $P$ relative to the LO
result, the latter operator can have a noticeable impact on the
quadrupole moment of deuterium, since the numerical value of $Q_d$
corresponds to a distance that is small on the typical scale of
deuteron physics $\sim 2$ fm. The operator is~\cite{Ch99}:
\begin{equation}
\langle {\bf p}'|J_0^{(5)}({\bf q})|{\bf p} \rangle=|e|(1 + \tau_1 \cdot
\tau_2) \frac{4 \pi h_4}{M \Lambda_Q^4} \left(\sigma_1 \cdot
{\bf q} \sigma_2 \cdot {\bf q} - \frac{|{\bf q}|^2}{3} \sigma_1 \cdot \sigma_2\right),
\label{eq:E2SD}
\end{equation}
and is designed to be diagonal in two-body spin and isospin and
contribute only for $S=1$ and $T=0$ states. In the case of deuterium
it represents an E2 photon inducing a ${}^3$S$_1 \rightarrow
{}^3$S$_1$ transition. Such a transition is possible because the
photon interacts with the total spin of the two-nucleon system through
the two-body operator (\ref{eq:E2SD}). The two-nucleon operator
(\ref{eq:E2SD}) will be induced when high-momentum modes in the $NN$
system are integrated out to obtain the low-momentum effective
theory. It could also be induced when heavy mesons which can couple to
a quadrupole photon in the requisite way are integrated out of the
$\chi$ET. This heavy-meson origin for the operator leads us to
anticipate that with a scale $\Lambda_Q$ of about 1.2 GeV the coupling
$h_4$ will be of order 1. In particular, if we used resonance
saturation in the $NN$ system~\cite{Ep01} to estimate the size of this
operator the first mesonic current that would contribute to the
operator would be the $\rho a_1 \gamma$ current~\cite{Tr01}. 

We now give arguments which demonstrate that physics at roughly this
scale could indeed remedy the discrepancy between the experimental
$Q_d$ and the result found from the mechanisms already discussed.  Let
us take the accepted number from ``high-quality'' potential models
$Q_d^{(0)}=0.270$ fm$^2$ (see, e.g.~Ref.~\cite{Wi95}). Calculations
with $\chi$PT wave functions obtain similar, or even slightly smaller,
numbers~\cite{Ph03,WM01,PC99,Ep05}.  Then, we adopt $\Delta
Q_d^{(3)}=0.008$ fm$^2$ as an estimate for the NLO and NNLO
corrections (which come mainly from the two-body operator
$J_0^{(3)}$). This leaves us with a remaining discrepancy between
theory and experiment of 0.008 fm$^2$, or about 3\%. Inserting the
operator (\ref{eq:E2SD}) between deuteron wave functions we obtain its
contribution to $Q_d$ as:
\begin{equation}
\Delta Q_d^{\rm SD}=\frac{32 \pi h_4}{M \Lambda_Q^4} |\psi(0)|^2,
\end{equation}
where $\psi(0)$ is the deuteron wave function at the origin. While
$\psi(0)$ is not an observable, neither is the dimensionless number
$h_4$: it is a wave-function-regularization dependent
coefficient. Since only S-waves contribute at $r=0$ if we assume
$\Lambda_Q=1.2$ GeV we can constrain the combination of $\psi(0)$ and
$h_4$ to be:
\begin{equation}
h_4 \left[\lim_{r \rightarrow 0} \frac{u(r)}{r}\right]^2 \approx 6.5~{\rm fm}^{-3},
\end{equation}
where $u(r)$ is the ${}^3$S$_1$ radial wave function.
Therefore the operator (\ref{eq:E2SD}) can be associated with physics
at scales of about 1 GeV and still remedy the discrepancy between the
theoretical value of $Q_d$ (including the
meson-exchange contribution to the charge (\ref{eq:J02Bmutilde}))
and the experimental value (\ref{eq:Qdexpt}).

This higher-order effect has an importance greater than one would
anticipate in Weinberg's counting (\ref{eq:sum}) not because it is
unnaturally large, but because, from the point of view of that
counting, $Q_d$ is unnaturally small~\cite{PC99}. This is hardly a surprise,
since, at both leading and next-to-leading order, $Q_d$ is generated
by one-body operators that connect the deuteron S-state to the
deuteron D-state. Such effects are suppressed by the
ratio $\eta=A_D/A_S=0.0253(2)$~\cite{St95}. In contrast, the operator
(\ref{eq:E2SD}) is not $\eta$-suppressed and so its contribution to
$Q_d$ is significantly larger than the naive estimate of $P^5 \sim
0.1$\% leads us to anticipate. In this context it is worth noting
that such an estimate for the short-distance effects in $\langle r_d^2
\rangle$ is completely validated by calculation~\cite{Ph03}. The
leading-order piece of $\langle r_d^2 \rangle$ is of the expected size
$\sim 1/m_\pi^2$, and two-body contributions, beginning with
effects from $J_0^{(3)}$ and continuing through the two-pion-exchange
mechanisms of $J_0^{(4)}$ and the C0-photon short-distance operator of
$J_0^{(5)}$, give contributions of the expected size, 
approximately $0.3$\%.

While this is reassuring, the (relatively) large impact of $J_0^{(5)}$
on $G_Q(0)$ means that we must ask how well we know this operator. Its
value at $Q^2=0$ can be fixed by the requirement that it repair the
discrepancy between theory and experiment for $Q_d$. But at this stage
we know nothing about its $Q^2$ dependence. For the purposes of this
paper we will assume that this operator arises from heavy-meson
exchange, and so model its $Q^2$ dependence by:
\begin{equation}
\Delta G_Q^{(5)}=\frac{\Delta Q_d^{\rm SD}}{\left(1 + \frac{{\bf
      q}^2}{\Lambda^2}\right)^5}.
\label{eq:J05uncert}
\end{equation}
The uncertainty in the operator is now encoded as uncertainty in the
scale $\Lambda$ of its momentum variation. We anticipate $\Lambda \geq
1.2$ GeV, because there are no meson resonances below 1.2 GeV which,
when integrated out of the theory, will yield this operator.  The only
danger with this reasoning is that two-pion-range mechanisms that occur
at $O(e P^5)$ may ultimately prove to be responsible for the $Q_d$
discrepancy. This possibility is under
investigation~\cite{quadri}. However, evaluation of relevant processes
in models which calculate, e.g. the role of $\Delta \Delta$ components
in deuterium, suggest that the dominant two-pion-exchange
contributions to $J_0^{(5)}$ are not large enough to remedy the $Q_d$
discrepancy~\cite{Bl89,AD98}. 

As for an upper bound on the value $\Lambda$; the effects of the
operator (\ref{eq:J05uncert}) persist to higher $Q^2$ as the scale of
its momentum variation is increased. At the $Q^2$'s considered here,
the impact of this operator on observables when we choose $\Lambda=2$
GeV is within 30\% of what one would obtain at $\Lambda=\infty$, so we
will vary $\Lambda$ between 1.2 and 2 GeV in order to assess the
theoretical uncertainty of our calculation. We will see below that
even with this range of variation our ignorance as
to the precise value of $\Lambda$ (or the precise function of $Q^2$
that modulates the current $J_0^{(5)}$) is the dominant contribution to our
theoretical uncertainty in the ratio $G_C/G_Q$.

Our goal in introducing the $Q^2$-dependence (\ref{eq:J05uncert})
into our calculation is to assess the potential impact on
our $\chi$ET calculation from physics that is not explicitly included
in it. The $Q^2$-dependence of $J_0^{(3)}$ will be modified by these
sorts of effects, as well as by higher-order loop contributions that
can be calculated in the $\chi$ET. However, once such higher-order
calculations are carried out the $Q^2$-dependence of $J_0^{(3)}$ can
presumably be constrained by input from electro-production in the
single-nucleon sector. Therefore here we take the operator $J_0^{(3)}$
as given. When we quote ranges for its impact on observables those
ranges arise from the fact that $\langle M'|J_0^{(3)}|M \rangle$ varies
when evaluated with different deuteron wave functions.

\section{Results for $G_C$ and $G_Q$}

\label{sec-J0results}

In this section we present results for the matrix elements of the
deuteron charge operator: $G_C$ and $G_Q$.

\begin{table}[htbp]
\begin{center}
\begin{tabular}{|c|c|c|c|}
\hline
Wave function & Order  &$\bar{\Lambda}$ (MeV) & $\Lambda$ (MeV)\\
\hline\hline%
001 & NLO  & 400  & 500 \\
002 & NLO &  550 & 500 \\
003 & NLO & 550 & 600 \\
004 & NLO & 400 & 700 \\
005 & NLO & 550 & 700 \\
\hline
101 & NNLO  & 450  & 500 \\
102 & NNLO & 600 & 500 \\
103 & NNLO & 550 & 600 \\
104 & NNLO & 450 & 700 \\
105 & NNLO & 600 & 700 \\
\hline
221 & N$^3$LO  & 450  & 500 \\
222 & N$^3$LO & 600 & 600 \\
223 & N$^3$LO & 550 & 600 \\
224 & N$^3$LO & 450 & 700 \\
225 & N$^3$LO & 600 & 700 \\
\hline\hline
\end{tabular}
\end{center}
\caption{\label{table-evgenywfs} Values of the SFR cutoff
  $\bar{\Lambda}$ and the Lippmann-Schwinger equation cutoff $\Lambda$
  that are employed in the different wave functions of
  Ref.~\cite{Ep05} that are used to compute deuteron form factors in
  this work. The wave functions are in groups of five: first those
  generated with the NLO $\chi$ET potential, then
  NNLO, then N$^3$LO.}
\end{table}

\begin{figure}[htb]
\centerline{\includegraphics*[width=110mm]{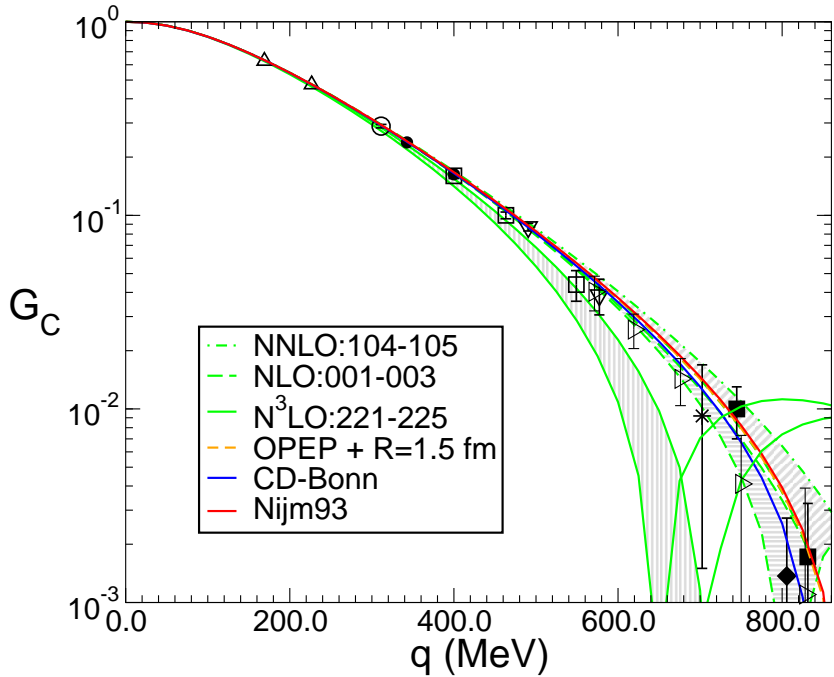}}
\caption{Results for the charge form factor $G_C$ as a function of
  $|{\bf q}|\equiv \sqrt{Q^2}$. The dashed lines show the largest and
  smallest form factors obtained with the NLO wave functions of
  Ref.~\cite{Ep05}. The range of predictions with these wave functions
  is given by the horizontally shaded region.  Similarly for the
  dot-dashed lines and the diagonally shaded region at NNLO, and the
  solid lines and the vertically shaded region at N$^3$LO. Other
  results shown are for a wave function from Ref.~\cite{PC99} with
  $R=1.5$ fm (short-dashed line), and the Nijm93 and CD-Bonn deuteron
  wave functions (solid red and blue lines respectively).  The
  experimental data is taken from the extraction of Ref.~\cite{Ab00B}
  and from Ref.~\cite{Ni03}:
  upward triangles represent data from the $T_{20}$ measurement of
  Ref.~\cite{Dm85}, open circle \cite{Fe96}, solid circle \cite{Sc84},
  open squares \cite{Bo99}, downward triangles \cite{Gi90}, rightward
  triangles~\cite{Ni03}, star \cite{Bo91}, solid squares \cite{Ga94},
  solid diamonds \cite{Ab00A}.}
\label{fig-GC}
\end{figure}

In Figure~\ref{fig-GC} we show the results for $G_C$ when the NNLO
[$O(e P^3)$] operator for $J_0$ is used (with nucleon structure
included via Eq.~(\ref{eq:factor})). The constants employed in
evaluating the operator were $g_A=1.29$, $f_\pi=93.0$ MeV, $m_\pi=139$
MeV, and $M=938.9$ MeV.  The dashed, dot-dashed, and solid lines show
the range of predictions generated using NLO, NNLO, and N$^3$LO wave
functions with different regulator masses $\Lambda$ and
$\bar{\Lambda}$. A list of the values of $\bar{\Lambda}$ and $\Lambda$
that are chosen for the wave functions used here is given in
Table~\ref{table-evgenywfs} (which is adapted from
Ref.~\cite{Ep05}). At a given order in the expansion for the chiral
potential these wave functions all have the same long-distance part,
but the different scales at which spectral-function regularization is
applied to obtain the $NN$ potential, and at which the
$\exp(-p^6/\Lambda^6)$ regulator is applied to the potential before
its insertion into the Lippmann-Schwinger equation, mean that they
differ in their short-distance physics. Therefore the amount by which
predictions for $G_C$ vary once the order of the wave-function
calculation is fixed gives us a lower bound for the impact of
short-distance physics on our calculation.

It is worth noting that the NLO potential only includes $O(P^2)$
corrections to $V$, and so the calculations labeled ``NLO'' in
Fig.~\ref{fig-GC} are limited by the accuracy of the $NN$
potential. The calculations labeled ``NNLO'' have $O(P^3)$
corrections included in the potential: the same level of accuracy to
which the operator $J_0$ has been computed. In this respect only the
computation with the NNLO wave functions is one that is carried out to
a consistent order in both the wave functions and operators obtained
using $\chi$ET. The results with the NLO and N$^3$LO wave functions
are shown for comparison. For the same reason we show results with the
Nijm93 wave function, the CD-Bonn wave function, and the
one-pion-exchange plus square well wave functions of
Ref.~\cite{PC99}. In each case consistent choices for $\beta$ and
$\nu$ (as explained in Sec.~\ref{sec-oneoverM}) were employed. In the
case of all but the CD-Bonn and $\chi$ET N$^3$LO calculations the
impact of the $p^2/M^2$ pieces of the one-pion-exchange potential and
the kinetic-energy operator has been assessed via the techniques
described in Appendix~\ref{ap-p2overM2corrns}.  Therefore all of these
matrix elements have the same one-pion-exchange physics. A difference
in their predictions is then either due to physics at the range of
two-pion-exchange, or to physics at distances less than the scale
where the chiral expansion can be used to reliably calculate the $NN$
potential.

In fact, Fig.~\ref{fig-GC} shows us that all of these wave functions
predict very similar form factors for $|{\bf q}| \leq 600$ MeV.  The most
noticeable difference occurs around the zero of $G_C$---a region
where, by definition, higher-order contributions cancel with
lower-order contributions, and the calculation is therefore sensitive
to the addition of higher-order effects. Most wave functions also
produce a $G_C$ in good agreement with the data compilation of Abbott
{\it et al.}~\cite{Ab00B} for $|{\bf q}| \leq 600$ MeV. An exception is the
predictions using the N$^3$LO wave function of Ref.~\cite{Ep05}, which
diverge from those of the other wave functions considered here at
significantly lower $Q^2$. Note also that the N$^3$LO predictions seem
to be significantly more sensitive to short-distance physics than is
the case for the wave functions computed with NLO or NNLO chiral
potentials, or even than the wave functions of Ref.~\cite{PC99}, where
only the result with the square well and surface delta function with
$R=1.5$ fm is shown, but changing $R$ to 2.5 fm produces a barely
discernible change in the short-dashed line.

It is possible that the situation with the predictions from the
N$^3$LO potential will improve when the pieces of $J_0$ of $O(e P^4)$
and $O(e P^5)$ which are not included in this calculation are added to
the current operator. But, even if this is the case, sizeable
cancellations between lower and higher-order effects are 
necessary if the N$^3$LO wave function is to be used to describe
electron-deuteron data at momentum transfers $Q^2 \geq 0.2$ GeV$^2$.

One might argue that one does not expect the $\chi$ET to work beyond
this scale anyway. However, the chiral expansion developed here and in
Refs.~\cite{PC99,WM01,Ph03} for the deuteron current operator still converges
well at $Q^2 \sim 0.3$ GeV$^2$. In part this is because
the impulse (leading-order) result for $G_C$ can be written as:
\begin{equation}
G_C^{(0)}({\bf q})=\int_0^\infty \,  d\hbox{r} \, j_0\left(\frac{|{\bf q}| r}{2}\right)
(u^2(r) + w^2(r)),
\label{eq:GCcoord}
\end{equation}
where $j_0$ is a spherical Bessel function and $w(r)$ is the
${}^3$D$_1$ radial wave function.  This means that---at least for the
impulse-approximation piece of the matrix element---the relevant
momentum scale at which the deuteron wave function is probed is not,
in fact, $|{\bf q}|$, but $|{\bf q}|/2$---half of the momentum
transfer is taken away by the center-of-mass degree of freedom. In
this context the failure of the N$^3$LO wave functions' form-factor
predictions to agree with the data when $|{\bf q}|/2 \approx 200$ MeV is
rather disturbing.

\begin{figure}[htbp]
\centerline{\includegraphics*[width=120mm]{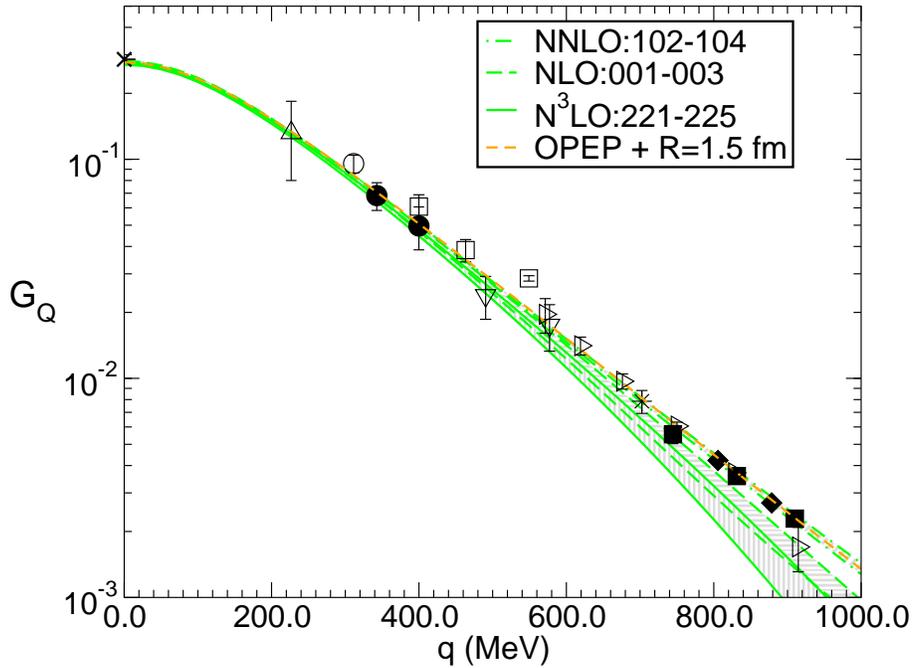}}
\caption{Results for the charge form factor $G_Q$ (in units of fm$^2$)
  as a function of $|{\bf q}|\equiv \sqrt{Q^2}$. Legend as in
  Fig.~\ref{fig-GC}, apart from the absence of curves for the Nijm93
  and CD-Bonn wave functions. These curves fall within the range of
  the dot-dashed lines, i.e. the results with NNLO $\chi$ET wave
  functions.}
\label{fig-GQ}
\end{figure}

In Fig.~\ref{fig-GQ} we show the results for $G_Q$. As mentioned
above, the shape produced by all wave functions is remarkably
similar---a point which was exploited in an extraction of $G_E^{(n)}$
in Ref.~\cite{SS01}. This can be understood from the presence of
$j_2$, instead of the $j_0$ of Eq.~(\ref{eq:GCcoord}) in the
co-ordinate space integral that gives the leading-order contribution
to $G_Q$. The pattern of convergence for $G_Q$ predictions with the
order of the $\chi$ET potential is interesting. The NLO band is quite
wide, and its centroid is below the data. The NNLO band is very
narrow, and its centroid agrees well with data out to 800 MeV. We
stress that this is the consistent order for computation of the
potential, given the current operator we have used. The N$^3$LO band
is then as wide, and below, the NLO band. In this plot predictions
using the Nijm93 and CD-Bonn potentials are not shown. But they lie
within the diagonally-shaded band that represents the range of
predictions of the NNLO $\chi$ET potential.

\begin{table}[htbp]
\begin{center}
\vskip 0.6 true cm
\begin{tabular}{||l||c|c|c|c|c|c||}
\hline \hline
 & Experiment & NLO:001--003 & NNLO:104--102 & N$^3$LO:221--225 & Nijm93\\
\hline  \hline
$Q_d$ (fm$^2$)     & 0.2859(3)  & 0.278--0.280   & 0.279--0.282  & 0.270--0.274   &0.276
\\ \hline \hline  
\end{tabular}
\end{center}
\caption{Deuteron quadrupole moment computed with our NNLO charge
  operator and different wave functions. Results are accurate to the
  number of digits shown. The ranges are generated by considering
  various values of $\Lambda$ and $\bar{\Lambda}$ at a given order in
  the expansion for the $\chi$ET $NN$ potential. The ``extremal'' wave
functions are indicated in the top line of the table.}
\label{table-Qd}
\end{table}

\begin{figure}[phtb]
\centerline{\includegraphics*[width=87mm]{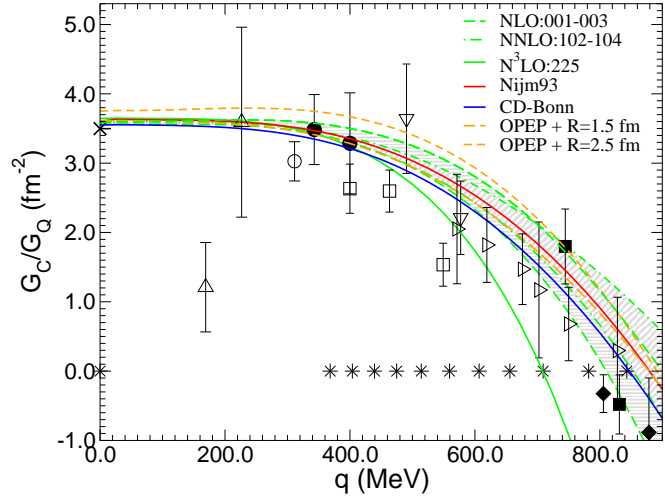}}
\caption{Results for the ratio $G_C/G_Q$. As in Fig.~\ref{fig-GC} the
  dashed lines and the horizontally shaded region show the range of
  results obtained with the NLO wave functions of
  Ref.~\cite{Ep05}. Likewise for the dot-dashed lines and the
  diagonally shaded region at NNLO, and the solid lines and the
  vertically shaded region at N$^3$LO. Other results shown are for
  wave functions from Ref.~\cite{PC99} with $R=1.5$ fm and $R=2.5$ fm
  (short-dashed lines), and the Nijm93 and CD-Bonn deuteron wave
  functions (solid red and blue lines respectively).  Upward triangles
  are data from the $T_{20}$ measurement of Ref.~\cite{Dm85}, open
  circle \cite{Fe96}, solid circle \cite{Sc84}, open squares
  \cite{Bo99}, downward triangles \cite{Gi90}, rightward triangles
  \cite{Ni03}, star \cite{Bo91}, solid squares \cite{Ga94}, solid
  diamonds \cite{Ab00A}. The asterisks indicate the points where
  BLAST will extract this ratio from their $t_{20}$ data.}
\label{fig-GCoverGQnoCT}
\end{figure}

In order to remove the rapid fall-off in the plots of
Fig.~\ref{fig-GQ} and \ref{fig-GC}, and also provide a result for the
ratio of form factors which will be measured at BLAST at the momentum
transfers indicated by the asterisks, we show predictions for the
ratio $G_C/G_Q$ in Fig.~\ref{fig-GCoverGQnoCT}. These predictions are
compared to data from the compilation of Ref.~\cite{Ab00B}, as well as
the more recent data set from Novosibirsk~\cite{Ni03}~\footnote{The
  experimental error bars in this plot, and in the plots of $G_C/G_Q$
  below, were obtained by summing the relative
  errors for $G_C$ and $G_Q$ given in Refs.~\cite{Ab00B,Ni03}. Some of
  the measurements of $t_{20}$ and $T_{20}$ from which these data came
  were quite precise, and so this procedure may well overestimate the
  size of their errors.}.  As can easily be gleaned
from Fig.~\ref{fig-GCoverGQnoCT}, the different wave functions considered give
predictions that disagree at about the 5\% level at $Q^2=0$, i.e. they
give different numbers for the deuteron quadrupole moment $Q_d$.

Numerical results for $Q_d$, computed with the NNLO operator that was
used to generate the predictions of Fig.~\ref{fig-GCoverGQnoCT}, are
presented in Table~\ref{table-Qd}. Note that the variation in the
results with the $\chi$ET NLO and NNLO wave functions as the cutoffs
$\Lambda$ and $\bar{\Lambda}$ are varied is of the same size as the
discrepancy between those predictions and the experimental value
(\ref{eq:Qdexpt}).

\begin{figure}[phtb]
\centerline{\includegraphics*[width=87mm]{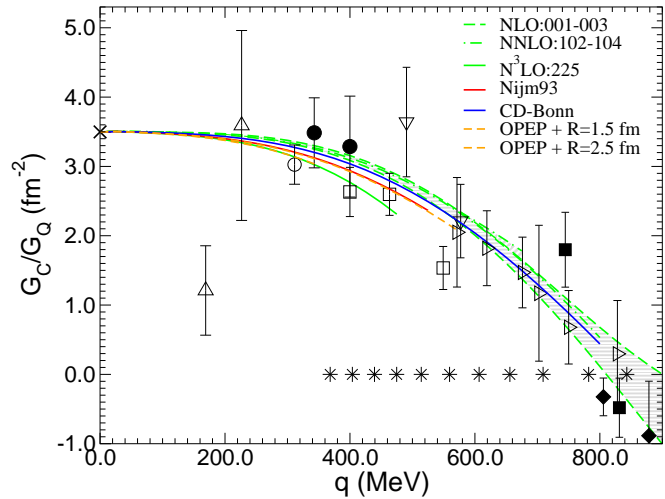}}
\caption{Results for $G_C/G_Q$, after $J_0^{(5)}$ is inserted with a
  coefficient adjusted to reproduce the experimental value of $Q_d$
  and $\Lambda=1.5$ GeV. Legend as in Fig.~\ref{fig-GCoverGQnoCT}. 
  Each curve is shown only up to the point where the $J_0^{(5)}$
  contribution is so large that the calculation is no longer
  reliable (with the exception of NLO).}
\label{fig-GCoverGQrenorm}
\end{figure}

The fact that short-distance physics can affect the value of $Q_d$ at
the 2--3\% level needed to restore agreement between theory and data
encourages us to incorporate the operator $J_0^{(5)}$ (see
Eq.~(\ref{eq:E2SD})) in our calculation. In doing so we adopt the
$Q^2$-dependence of Eq.~(\ref{eq:J05uncert}) with $\Lambda=1500$
MeV. For each wave function the value of $h_4$ is adjusted to yield
the experimental value of $Q_d$. The results of this procedure are
presented in Fig.~\ref{fig-GCoverGQrenorm}. The NNLO, N$^3$LO, Nijm93,
CD-Bonn, and one-pion-exchange-tail potential curves are only shown
out to the momentum transfer where the contribution of the operator
(\ref{eq:E2SD}) makes up 20\% of the contributions from the preceding
orders. This gives a way to estimate where the calculations with
various wave functions become unreliable: they are unreliable when
$J_0^{(5)}$ is no longer a small piece of the total $G_C/G_Q$.  Under
this criterion most of the wave functions can give reliable
predictions to $|{\bf q}|=500$--$600$ MeV, and below this value the
wave-function dependence is markedly reduced as compared to what is
seen in Fig.~\ref{fig-GCoverGQnoCT}.  Note that the lines at NLO are
shown only to give an idea of the uncertainty coming from sensitivity
to the choice of $(\Lambda,\bar{\Lambda})$. Their predictions with
this wave function for $G_C/G_Q$ beyond 600 MeV are provided only for
recreational purposes.

\begin{figure}[hbpt]
\centerline{\includegraphics*[width=90mm]{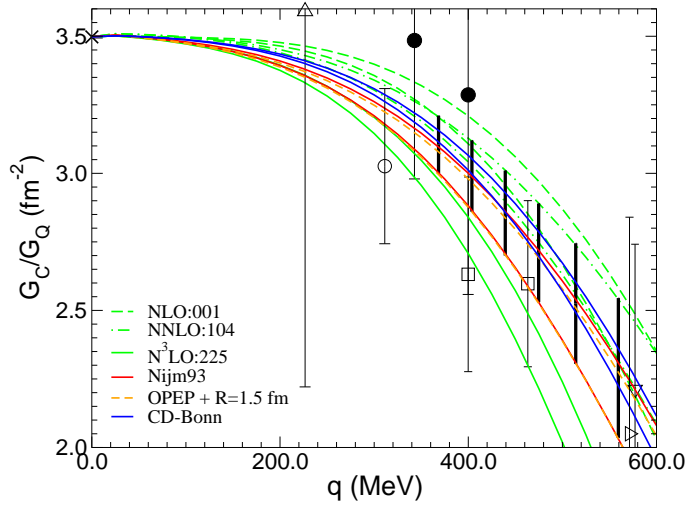}}
\caption{Results for $G_C/G_Q$, showing the variation that results
  from ignorance of the $Q^2$-dependence of the operator $J_0^{(5)}$
  ($\Lambda=1.2$--$2$ GeV).  Calculations are shown for NLO (long
  dashed), NNLO (dot-dashed), N$^3$LO (solid green), Nijm93 (solid
  red), CD-Bonn (solid blue) and $R=1.5$ fm square well + one-pion
  exchange (short dashed) wave functions. The vertical lines indicate
  a reasonable range for the theoretical prediction at each of the
  points where BLAST will have data. The experimental data is from
  Refs.~\cite{Dm85} (upward triangle), \cite{Fe96} (open circle),
  \cite{Sc84} (solid circle), and \cite{Bo99} (open square). Note
  change of scale as compared to Fig.~\ref{fig-GCoverGQrenorm}.}
\label{fig-GCGQdetail}
\end{figure}

\begin{table}[hbtp]
\begin{center}
\begin{tabular}{|c|c|c|c|}
\hline
$|{\bf q}|$ (MeV) & $G_C/G_Q$ (fm$^2$) & Error bar (fm$^2$)\\
\hline\hline%
368.4 & 3.11 & 0.11\\
403.9 & 2.99 & 0.13\\
439.3 & 2.86 & 0.16\\
474.8 & 2.71 & 0.18\\
514.2 & 2.53 & 0.22\\
559.5 & 2.29 & 0.26\\
606.8 & 2.02 & 0.30\\
\hline\hline
\end{tabular}
\end{center}
\caption{\label{table-BLASTpreds} 
Predictions for the ratio $G_C/G_Q$ at the values of $|{\bf q}|$ where
this quantity will be measured at BLAST. The error bar and central
values displayed here are obtained via the procedure discussed in the text.}
\end{table}

Lastly we focus on the region where the $\chi$ET is 
reliable: $|{\bf q}| \leq 600$ MeV. An enhanced view of this region is shown
in Fig.~\ref{fig-GCGQdetail}. For each choice of $NN$ potential two
curves are shown: the upper one of the pair corresponds to choosing
$\Lambda=1.2$ GeV when evaluating the operator $J_0^{(5)}$ and the
lower one corresponds to choosing $\Lambda=2$ GeV. A conservative
estimate for the impact of short-distance physics which is not
well-constrained in this $\chi$ET calculation is given by combining
the uncertainties from $(\Lambda,\bar{\Lambda})$ variation and the
uncertainty coming from lack of knowledge about the momentum
dependence of $J_0^{(5)}$. The black bars then represent the range of
possible values that the $\chi$ET predicts for $G_C/G_Q$ at the
kinematics where there will be data from BLAST. These ranges are
reproduced in Table~\ref{table-BLASTpreds}. The error is about $\pm
3$\% at the lowest BLAST point and increases with $Q^2$, as it
should. Note that we have not included the N$^3$LO $\chi$ET wave
function, or the NLO $\chi$ET wave function, in generating these
predictions. As already discussed, the predictions for $G_C$ and $G_Q$
with the N$^3$LO wave function deviate already from the extant data at
quite low $Q^2$, while the NLO wave function is, in the $\chi$ET sense,
less accurate than the operator being used here. The predictions
obtained with these wave functions are, however, within 2$\sigma$, if
the theoretical error bars we have obtained are taken to have the
usual one-standard-deviation interpretation.

\section{Results for $G_C/G_M$}

\label{sec-Jplusresults}

\begin{figure}[htb]
\centerline{\includegraphics*[width=110mm]{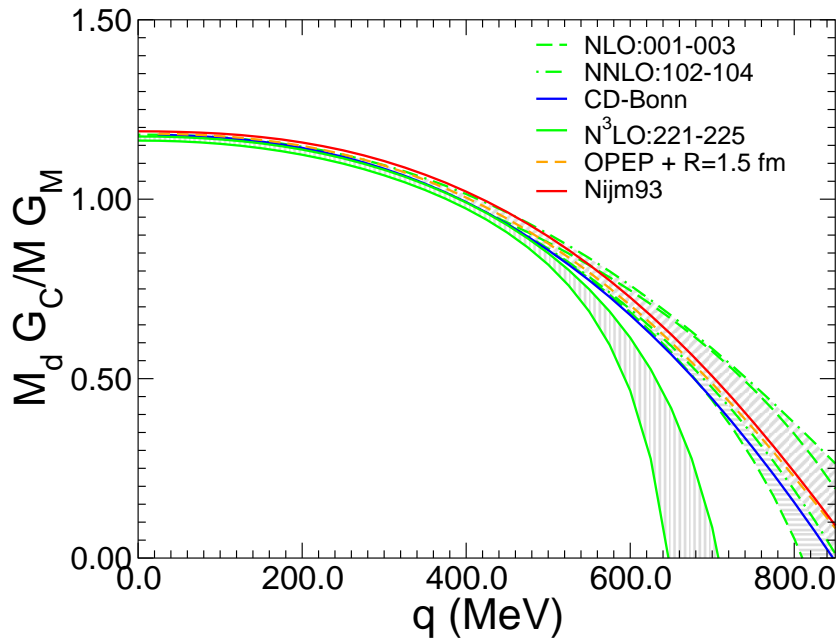}}
\caption{Results for the ratio $G_C/G_M$. Calculations shown are for
  extremal NLO (long dashed), extremal NNLO (dot-dashed), extremal
  N$^3$LO (solid green) (solid red), CD-Bonn (solid
  blue) and $R=1.5$ fm square well + one-pion exchange (short dashed)
  wave functions.}
\label{fig-GCoverGM}
\end{figure}

BLAST will also measure the ratio $G_C/G_M$. Predictions for that
observable are provided in Fig.~\ref{fig-GCoverGM}. We do not show any
experimental data in Fig.~\ref{fig-GCoverGM} because, as far as we can
glean from the literature, all previous data on $G_M$ comes from different data
sets to that used for the extraction of $G_Q$ and
$G_C$~\cite{Ab00B}. Therefore in general data on $G_C$ and $G_M$ are
not at the same $Q^2$ and have different systematic errors. The BLAST
data set will be pioneering in this regard. 

The calculations displayed in Fig.~\ref{fig-GCoverGM} are accurate to
relative order $P^2$, although the $G_C$ used here is actually
computed up to relative order $P^3$. Once again the shape of the
low-momentum part of the ratio is fairly wave-function independent,
but the value at $Q^2=0$ changes as we move through the different wave
functions used for the computation of Fig.~\ref{fig-GCoverGM}, due to
short-distance contributions to $\mu_d$ being different for different
wave functions. However, even without renormalization there is a
robust prediction for the ratio out to $Q^2 \approx 0.1$ GeV$^2$. The
robust prediction is that $G_C/G_M$ is approximately flat. This would
be exactly the case in the absence of relativistic, meson-exchange,
and nucleon-structure contributions to the operator, and if
$w(r)=0$. The relativistic corrections at $O(P^2)$ are negligible at
$Q^2 < 0.1$ GeV$^2$, and the meson-exchange piece of the charge
operator is higher order than we are attempting to calculate the ratio
$G_C/G_M$ to. As far as the operator is concerned this leaves only the
nucleon-structure effects, which depend on the ratio:
\begin{equation}
\frac{G_E^{(s)}}{G_M^{(s)}}.
\end{equation}
If a strict chiral expansion is used for the form factors then this
ratio depends (again, at this order) on the difference of the
isoscalar magnetic and charge radii, and amounts to a $< 10$\% effect at
$Q^2=0.1$ GeV$^2$. Even though taking the ratio $G_C/G_M$ does
not (as it did in the case of the ratio of the previous section)
eliminate nucleon-structure effects, it does reduce their
size. Meanwhile, the effects of $w$ grow with $Q^2$, and so at $Q^2 <
0.1$ GeV$^2$ it is thus not particularly surprising that $G_M/G_C$ is,
to quite a good approximation, flat.

Given the extent of the variation in the prediction for the ratio
beyond $|{\bf q}|=500$ MeV it is difficult to believe that the NLO
predictions for the ratio shown here are reliable beyond that
point. This situation could improve once the NNLO pieces of the
operator ${\bf J}$ were computed, but short-distance pieces of ${\bf
  J}$ appear already at that order. Therefore it is a prediction of
the chiral expansion that this ratio will be more sensitive to
short-distance physics than is $G_C/G_Q$. The position of the minimum
in $G_M$ is known to be very sensitive to such short-distance
physics~\cite{Wi95,vO95,SP02,Ph05B}. In this context it is worth
noting that the minimum for the N$^3$LO wave functions is already at
$|{\bf q}| \approx 800$ MeV---much lower than in any of the
calculations of Refs.~\cite{Wi95,vO95,SP02,Ph05B} and indeed, much
lower than the experimental data allows the position of the minimum to
be~\cite{Ab00B}.

\section{Conclusion}

\label{sec-conclusion}

We have used the $\chi$ET isoscalar charge operator in the
nucleon-nucleon sector computed to NNLO (including a consistent
treatment of the $1/M$ pieces of the charge), and the wave functions of
Ref.~\cite{Ep05}, to obtain the form-factor ratios $G_C/G_E^{(s)}$,
$G_Q/G_E^{(s)}$, $G_C/G_Q$.  These ratios test $\chi$ET's ability to
describe deuteron structure. We confirm and extend the finding of
Ref.~\cite{Ph03}, that the NLO and NNLO $\chi$ET wave functions,
combined with the NNLO $J_0$, yield results for these ratios that
agree---within the experimental uncertainties---with the extractions
of Ref.~\cite{Ab00B} for $Q^2 < 0.35$ GeV$^2$. In contrast, the
N$^3$LO wave function of Ref.~\cite{Ep05}, when used in conjunction with
the N$^2$LO charge operator, produces form factors that depart from the
data at $Q^2 \approx 0.2$ GeV$^2$.

In light of the upcoming release of data on $t_{20}$ from BLAST we
examined the ratio $G_C/G_Q$ in detail. We found variation in the
$\chi$ET predictions for this ratio at $Q^2=0$, and also found
that---even allowing for this variation---$\chi$ET is in disagreement
with the experimental value for this quantity. This phenomenon---the
``$Q_d$ problem''---is familiar in high-precision $NN$ potential
models with the modern value of the $\pi$NN coupling constant. In
$\chi$ET its solution arises naturally through a higher-order two-body
current that couples exclusively to quadrupole photons. We added this
operator to our analysis, and showed that when we do so the $\chi$ET
predictions for $G_C/G_Q$ (with the NNLO wave functions) fall within a
narrow band out to $Q^2 \approx 0.35$ GeV$^2$ (see
Fig.~\ref{fig-GCGQdetail}). We also performed the calculation with the
same charge operator and potential-model wave functions that have a
one-pion-exchange tail identical to that of the $\chi$ET
potential~\cite{St94,Ma01,PC99}. We found that these wave functions
make the band obtained at NNLO in $\chi$ET about a factor of two
wider. We conservatively adopt the full width of that band as 
representative of the theoretical uncertainty in our calculation.

Meanwhile, the $\chi$ET predictions for $G_C/G_M$, which will also be
measured at BLAST, are not reliable to as high a $Q^2$.  In saying
this, it should, in fairness, be pointed out that ${\bf J}$ has not
been computed to as high an order as $J_0$, and it could therefore be
that $G_C/G_M$ can also be well described to $Q^2=0.35$ GeV$^2$ in
$\chi$ET once $O(e P^4)$ corrections to $\bf J$ are included. This is
a topic for future work. Another topic for future work is the
inclusion of the $O(e P^4)$ pieces of $J_0$ that were already computed
in Ref.~\cite{Pa00} at $Q^2=0$ in the finite-$|{\bf q}|$
calculation~\cite{quadri}.  In addition, the operators and the
$\chi$ET wave functions used in this paper to make predictions for
the BLAST data can be further tested by comparing their predictions
with experimental results for deuteron
electro-disintegration---although this will require the computation of
the isovector pieces of the operators. 

Irrespective of such future efforts, one thing is already clear from
Fig.~\ref{fig-GCGQdetail}.  When the theoretical predictions for
$G_C/G_Q$ are renormalized in the manner we advocate here, the
theoretical uncertainty in $G_C/G_Q$ for $Q^2 \leq 0.3$ GeV$^2$ is
less than the uncertainty in the data. This makes the
low-$Q^2$ data from BLAST all the more crucial, since it will provide
an important test of $\chi$ET's ability to organize contributions to
deuteron observables, and its ability to use that organization to
provide estimates of the theoretical uncertainty in a given
calculation.

\section*{Acknowledgments}
I thank Michael Kohl and Chi Zhang for a number of conversations
regarding the BLAST data, and in particular for stimulating questions
regarding the theoretical uncertainty that arises from the
$Q^2$-dependence of the two-body current that renormalizes $Q_d$.  I
am also grateful to Richard Milner for inviting me to a BLAST workshop
in January 2005 where a number of the results in this paper were
presented in preliminary form. Thanks also to Evgeny Epelbaum for
supplying me with the wave functions of Ref.~\cite{Ep05}, and for his
comments on both my results and this manuscript. I am also grateful to
Lucas Platter for his careful reading of the manuscript and assistance
with the spelling of Dutch names.  This work was supported by the US
Department of Energy under grant DE-FG02-93ER40756.


\appendix

\section{Including $p^2/M^2$ corrections in ``non-relativistic'' wave functions}

\label{ap-p2overM2corrns}

One way to include relativistic corrections to the nucleon kinetic
energy operator in the $\chi$ET is to solve the
``relativistic Schr\"odinger equation''~\cite{Ep05}
\begin{equation}
2 \left[\sqrt{p^2 + M^2} - M\right] \psi_l^{sj}(p) + \int 
\frac{dp' {p'}^2}{(2 \pi)^3} \,
V_{ll'}^{sj} (p,p') \psi_{l'}^{sj} (p')=E \psi_l^{sj}(p),
\label{eq:relSE}
\end{equation}
with the (partial-wave decomposed) potential $V_{ll'}^{sj}(p,p')$ the
one that is obtained from the $\chi$ET using the
counting rules explained in Section~\ref{sec-intro}.
To facilitate computation of the $p^2/M^2$ (and beyond) corrections to
the nucleon kinetic energy operator we adopt the Kamada-Gl\"ockle
transformation~\cite{KG98} and define a new relative momentum 
$\tilde{p}$ such that
\begin{equation}
\frac{\tilde{p}^2}{2 M}=\sqrt{M^2 + p^2} - M.
\label{eq:KG}
\end{equation}
The inverse transformation is then:
\begin{equation}
p=\tilde{p}\sqrt{1 + \frac{\tilde{p}^2}{4 M^2}} \equiv \tilde{p}
g(\tilde{p}).
\end{equation}
As shown by Kamada and Gl\"ockle, any ``relativistic'' Schr\"odinger
equation that employs the kinetic energy operator with the
relativistic form on the right-hand side of Eq.~(\ref{eq:KG}) can be
recast as a non-relativistic Schr\"odinger equation, i.e.~\cite{KG98,Ep05}
\begin{equation}
\frac{\tilde{p}^2}{2M} \tilde{\psi}_l^{sj}(\tilde{p}) + \int
\frac{d\tilde{p}' \tilde{p}^{\prime \, 2}}{(2 \pi)^3} \, \, 
\tilde{V}_{ll'}^{sj}
(\tilde{p},\tilde{p}') \tilde{\psi}_{l'}^{sj} (\tilde{p}')=E
\tilde{\psi}_l^{sj}(\tilde{p}),
\end{equation}
where the potential $\tilde{V}_{ll'}^{sj}$ is obtained from
$V_{ll'}^{sj}$ via:
\begin{equation}
\tilde{V}_{ll'}^{sj}(\tilde{p},\tilde{p}')=h(\tilde{p}) V_{ll'}^{sj}(\tilde{p}
g(\tilde{p}),\tilde{p}' g(\tilde{p}')) h(\tilde{p}'),
\label{eq:tildeV}
\end{equation}
and 
\begin{equation}
\tilde{\psi}(\tilde{p})=h(\tilde{p}) \psi(\tilde{p} g(\tilde{p})).
\label{eq:wfreln}
\end{equation}
Here the factor $h$ is introduced to ensure that, as long as $\psi$
is normalized to one, $\tilde{\psi}$ is also normalized to
one. Calculation of the Jacobian associated with the transformation 
(\ref{eq:KG}) yields:
\begin{equation}
h(\tilde{p})=\sqrt{\left(1 + \frac{\tilde{p}^2}{2 M^2}\right)
  g(\tilde{p})}.
\end{equation}

We can also work this procedure in reverse, i.e. interpret solutions
of the non-relativistic Schr\"odinger equation as solutions to a
dynamical equation with a relativistic kinetic energy operator and a
relativistic potential. (Something similar is done to obtain the $NN$
c.m. frame Hamiltonian in approaches to electron-deuteron scattering
based on Relativistic Hamiltonian Dynamics, see, for example,
Ref.~\cite{Ch84}.).  Here we will make this interpretation for the
deuteron wave functions obtained with the NLO and NNLO chiral
potentials.  

Suppose that $\tilde{V}$ is the chiral potential computed
(at NLO or NNLO) {\it without} $p^2/M^2$ corrections, i.e.
\begin{equation}
\tilde{V}_{ll'}^{sj}(\tilde{p},\tilde{p'})=f(\tilde{p}) {\cal
  M}_{ll'}^{sj}(\tilde{p},\tilde{p}') f(\tilde{p}'),
\label{eq:tildeVchiral}
\end{equation}
where ${\cal M}$ is the (partial-wave decomposed) sum of
$NN$-irreducible diagrams that is computed at NLO and NNLO using
spectral-function regularization in Ref.~\cite{Ep05} and
$f(p)=\exp(-p^6/\Lambda^6)$ is the ``Lippmann-Schwinger equation
regulator'' used there. A potential $V$ that includes $p^2/M^2$
corrections and is associated with the same sum of Feynman diagrams
can be constructed by inverting Eq.~(\ref{eq:tildeV}).  We find that
this potential, which when inserted in the relativistic Schr\"odinger
equation (\ref{eq:relSE}) will be phase-equivalent to $\tilde{V}$ of
Eq.~(\ref{eq:tildeVchiral}), is:
\begin{equation}
V_{ll'}^{sj}(p,p')=\frac{f(p)}{h(p)} 
 {\cal  M}_{ll'}^{sj} \left(\sqrt{2M \sqrt{M^2 + p^2} - 2M^2},\sqrt{2M \sqrt{M^2
    + p'^2} - 2M^2}\right) \frac{f(p')}{h(p')}.
\end{equation}
We have not bothered to distinguish between $\tilde{p}$ and $p$ in the
functions $f$ and $h$, because we now
define a new regulator function $\overline{f} \equiv f/h$ to obtain:
\begin{equation}
V_{ll'}^{sj}(p,p')=\overline{f}(p)
 {\cal  M}_{ll'}^{sj} \left(\sqrt{2M \sqrt{M^2 + p^2} - 2M^2},\sqrt{2M \sqrt{M^2
    + p'^2} - 2M^2}\right) \overline{f}(p'),
\label{eq:Vrearranged}
\end{equation}
up to terms that are of order $p^4/M^4$.  Since we count $M$ as the
same order as the cutoff scale $\Lambda$ absorbing the functions $h$
into the definition of the regulator in this way makes physical
sense. Apart from these short-distance effects the only difference
between the potentials $V$ and $\tilde{V}$ is the use of the
``stretched'' momenta in $V$.

If we wish we can also incorporate the factor $\left(1 - \frac{p^2 +
  p'^2}{2 M^2}\right)$ that distinguishes Eq.~(\ref{eq:relOPE}) from
the non-relativistic treatment of one-pion exchange used at NLO and
NNLO in Ref.~\cite{Ep05}. To do this we redefine the regulator again,
writing
\begin{eqnarray}
V_{ll'}^{sj}(p,p')=\overline{\overline{f}}(p) \left(1 - \frac{p^2}{2 M^2}\right)
 {\cal  M}_{ll'}^{sj}\left(\sqrt{2M \sqrt{M^2 + p^2} - 2M^2},\sqrt{2M
   \sqrt{M^2 + 
p'^2} - 2 M^2}\right)\nonumber&&\\
\left(1 - \frac{p'^2}{2 M^2}\right)
\overline{\overline{f}}(p'),&&
\label{eq:Vfinal}
\end{eqnarray}
where, up to the order to which we work:
\begin{equation}
  \overline{\overline{f}}(p)=\left(1 + \frac{3 p^2}{16 M^2}\right)
\exp\left(-\frac{p^6}{\Lambda^6}\right).
\end{equation}
If one ignores the regulator $\overline{\overline{f}}$, then the
factors in front of the one-pion-exchange part of the potential in
Eq.~(\ref{eq:Vfinal}) agree with Eq.~(\ref{eq:relOPE}) and the
expression for the two-pion-exchange pieces has been modified only at
N$^3$LO and beyond. As for the short-distance pieces of the potential,
the changes in the way the chiral $NN$ potential is regularized ($f
\rightarrow \overline{f} \rightarrow \overline{\overline{f}}$) mean
that the $NN$ LEC $C_2$ will have to be re-defined: its value is
shifted by a term of $O(1/M^2)$. However, $C_2$ is fitted to data, so
no change in observables should result from doing this. 

Therefore we have shown that, via modifications of the (unobservable)
short-distance part of the potential that do not affect the low-energy
observables, we can absorb much of the ``relativistic effects'' that
would enter the NLO and NNLO chiral potentials were we to count $M
\sim \Lambda$. The only such effect that cannot be absorbed by a
redefinition of the regulator used when solving the Lippmann-Schwinger
equation with the potential $V$ is the different arguments at which
${\cal M}$ is evaluated in Eq.~(\ref{eq:Vfinal}) as compared to
Eq.~(\ref{eq:tildeVchiral}). This difference is an NLO 
effect for the part of $V$ coming from one-pion exchange, where it
represents the interplay of the relativistic kinetic energy operator
and the (leading-order) one-pion-exchange potential. This interplay is
not included in our calculations, but all other relativistic effects
are, and as reported in Sec.~\ref{sec-oneoverM}, they prove to be small.

The main difference between interpreting the expression (\ref{eq:tildeVchiral})
as a non-relativistic potential and interpreting it as resulting from
a relativistic calculation after application of the Kamada-Gl\"ockle
transformation therefore resides in the momentum arguments that enter
the Schr\"odinger equation. In the first case the wave function that
is the solution of that equation will be a function of $\tilde{p}$.
In the second case it will be a function of
$p=\sqrt{2M \sqrt{M^2 + \tilde{p}^2} - 2 M^2}$. The relationship
between the two wave functions is given by Eq.~(\ref{eq:wfreln}). This can be
inverted to yield Eq.~(\ref{eq:wfreln2}), which we rewrite here:
\begin{equation}
\psi(p)=\sqrt{\frac{M}{\sqrt{M^2 + p^2}}} \left(\frac{2M}{M + \sqrt{M^2
      + p^2}}\right)^{1/4}
\tilde{\psi}\left(\sqrt{2M \sqrt{M^2 + p^2} - 2M^2}\right).
\end{equation}
Thus, if we are provided with the wave-function $\tilde{\psi}$ that is
obtained by solving the non-relativistic Schr\"odinger equation with
the potential $\tilde{V}$, then we may derive from that the wave
function $\psi$, that is the solution of the relativistic
Schr\"odinger equation (\ref{eq:relSE}) with a phase-equivalent
relativistic potential $V$ (\ref{eq:Vfinal}), with $V$ constructed to
include all the corrections to one-pion-exchange that must be present
in the NLO and NNLO potentials if we count $M \sim \Lambda$. This
procedure is not necessary in the case of the N$^3$LO potential. In
that case the $1/M^2$ corrections which are connected with the current
(\ref{eq:J02Bmutilde}) were explicitly computed in Ref.~\cite{Ep05}.

\end{document}